# *Molecular streaming and its voltage control in ångström scale channels*

T. Mouterde[1,†], A. Keerthi[2,3,†], A. R. Poggioli[1], S. A. Dar[2,3,4], A. Siria[1],
A. K. Geim[2,3,*], L. Bocquet[1,*] and B. Radha[2,3,*].

1. Laboratoire de Physique Statistique - PSL, École Normale Supérieure, UMR 8550 du CNRS
24 rue Lhomond, 75005 Paris, France

2. School of Physics and Astronomy, University of Manchester
Manchester, M13 9PL, United Kingdom.

3. National Graphene Institute, University of Manchester
Manchester, M13 9PL, United Kingdom.

4. Department of Physics, University of Engineering & Technology
KSK campus, Lahore, Pakistan

[†] these authors contributed equally to the work
* lyderic.bocquet@ens.fr, geim@manchester.ac.uk, radha.boya@manchester.ac.uk

**The field of nanofluidics has shown considerable progress over the past decade thanks to key instrumental advances, leading to the discovery of a number of exotic transport phenomena for fluids and ions under extreme confinement. Recently, van der Waals assembly of 2D materials[1] allowed fabrication of artificial channels with ångström-scale precision[2]. This ultimate confinement to the true molecular scale revealed unforeseen behaviour for both mass[2] and ionic[3] transport. In this work, we explore pressure-driven streaming in such molecular-size slits and report a new electro-hydrodynamic effect under coupled pressure and electric force. It takes the form of a transistor-like response of the pressure induced ionic streaming: an applied bias of a fraction of a volt results in an enhancement of the streaming mobility by up to 20 times. The gating effect is observed with both graphite and boron nitride channels but exhibits marked material-dependent features. Our observations are rationalized by a theoretical framework for the flow dynamics, including the frictional interaction of water, ions and the confining surfaces as a key ingredient. The material dependence of the voltage modulation can be traced back to a contrasting molecular friction on graphene and boron nitride. The highly nonlinear transport under molecular-scale confinement offers new routes to actively control molecular and ion transport and design elementary building blocks for artificial ionic machinery, such as ion pumps. Furthermore, it provides a versatile platform to explore electro-mechanical couplings potentially at play in recently discovered mechanosensitive ionic channels[4].**

Reducing the dimensions of fluidic devices down to the nanometre scale (e.g. nanotubes[5–9] or nanopores[10–15]) laid foundations for the discovery of new water and ion transport properties[16–18]. Recent advances in fabrication of capillaries using van der Waals assembly[1] of 2D materials allowed to push beyond the nanometric limit to ångström dimensions. The latter are not merely a symbolic limit but rather signify the breach into the molecular scale associated with the expected breakdown of continuum transport equations and, in particular, of the celebrated Navier-Stokes equation for viscous fluid transport[19]. Furthermore, below a lateral confinement of one nanometre, water rearranges into one or two layers resulting in strong suppression of its dielectric permittivity[20,21] and the possible formation of room-temperature ice[22]. Moreover, the ion hydrated diameter becomes comparable to the channels size so that hydration shells come into direct contact with the channel walls impeding ionic motion[3]. Under extreme confinement, surface effects inherently dominate, and water transport becomes strongly dependent on channel walls' material[23]. Below we explore the combined pressure and electric-field driven transport in sub-nanometre slit-like channels: this allows, for the first time, to probe in a fully controlled way with a known driving force, the impact of water flow on ionic transport at this ultimate scale, providing unique insights on molecular transport.

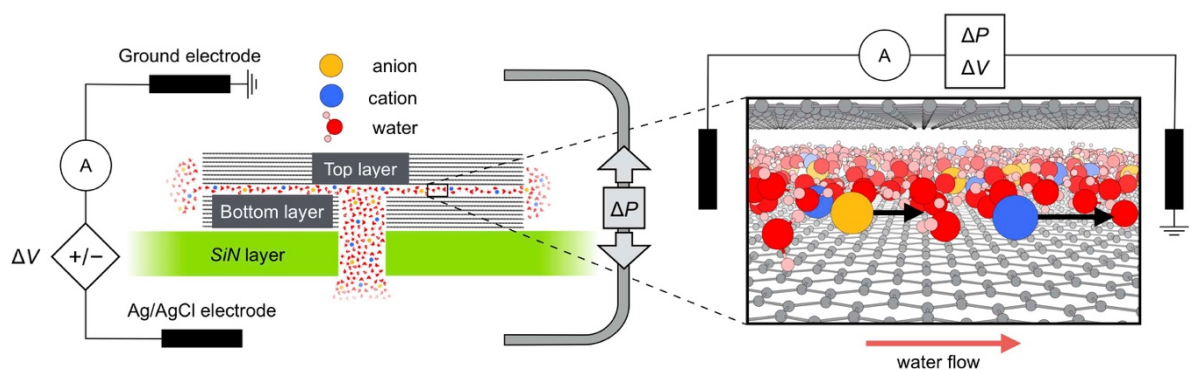

**Fig. 1 | Experimental setup for pressure and voltage driven current.** Its schematic: Ångström channels (fabricated on a Si/SiN wafer) separate two reservoirs containing KCl solutions. The entry and exit of the channel are on either side of the wafer. We set the voltage $\Delta V$ and the pressure $\Delta P$ along the channels and monitored the resulting current $I$. Right panel: Illustration of ions moving in water under strong confinement (only one layer of top and bottom graphite walls is shown for clarity). From the streaming current measurements, positive streaming currents indicate that potassium ions move faster than chloride ions inside the channel.

Our devices (Fig. 1) were ångström-scale channels fabricated on a Si/SiN substrate using the technology described in previous studies[2,3]. Briefly, the channels were made by van der Waals assembly of two (~10 nm and 150 nm) thin crystals of graphite separated by strips of bilayer graphene. Each device had $N = 200$ channels of height $h_0 \approx 6.8$ Å, width $w = 130$ nm and length $L$ of few micrometres (see Methods section 'Device fabrication' and Extended Data Fig. 1).

The channels were assembled on top of a micrometre opening etched in the Si/SiN wafer, which defined the entry of the fluidic channels whereas the exit was on the other side of the wafer (Fig. 1). The channels connect two macroscopic reservoirs filled with KCl solutions of concentration $c$. The electrical current $I$ was measured using chlorinated Ag/AgCl electrodes placed in the reservoir. The net current is typically on the order of a few pA per channel (for few tens of mV) at high salt concentration as previously reported[3]. In the concentration range that has been explored here the current varies linearly with voltage and concentration, in agreement with the previously reported small surface charge for this ultra-confined system[3]. Below we focus on the ionic current driven by the pressure drop $\Delta P$ applied across the channels and, also, explore the effect of the additional potential difference $\Delta V$ along the channel. $\Delta V$ was controlled by a patch-clamp amplifier (ground electrode is on the top side) to achieve a resolution in tenth of pA whereas the pressure was provided by a pump connected to the reservoir (Methods section 'Streaming current measurements'). We could apply the pressure in both directions and found no influence on the reported results (see Extended Data Fig. 2). The pressure applied from the bottom side in Fig. 1 is denoted as positive. We also performed control experiments with similar devices but without channels. Application of $\Delta P$ or $\Delta V$ in the control case did not result in any current, confirming that our devices were structurally stable and, for example, did not delaminate under pressure (see Methods section 'Streaming current measurements' and Extended Data Fig. 3).

This setup (Fig. 1) allows us to measure the pressure driven component of the ionic current, referred to as streaming current, $I_{str} = I(\Delta P, \Delta V) - I(0, \Delta V)$. The streaming current provides an indirect measure of the water flow under the confinement. Fig. 2a shows the time response of $I_{str}$ if applying $\Delta P$ up to 125 mbar in 25 mbar increments. Each step lasts 20 s, and the delay between successive steps is 20 s. After an initial overshoot, $I_{str}$ rapidly reaches a steady state and, once the pressure is released, quickly returns to zero. Furthermore, the current is positive for positive applied pressures, which corresponds to a flow conveying a net positive charge that gradually increases with the pressure gradient, $\Delta P/L$. This is consistent with the reduction in chloride mobility as compared to that of potassium under strong confinement[3] (right panel of Fig. 1).

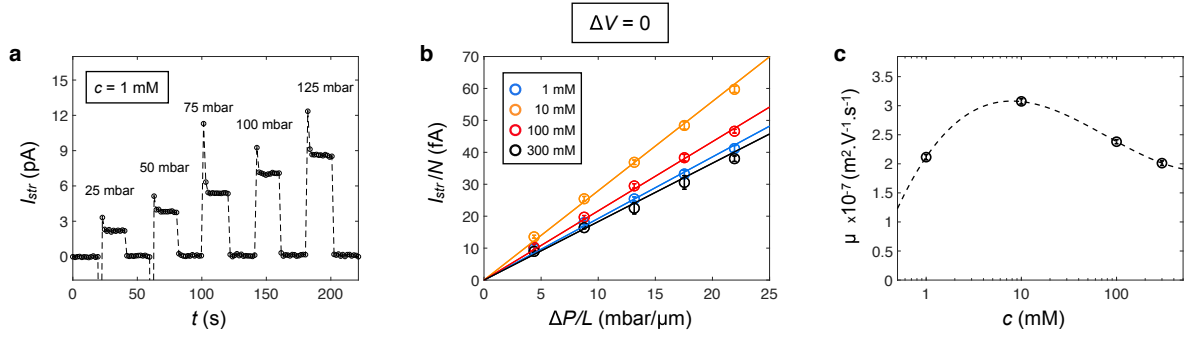

**Fig. 2 | Pressure-driven current without applying bias. a,** $I_{str}$ as a function of time for graphite channels; $c = 1$ mM; $L = 5.7 \pm 0.1$ μm. Current overshoots once the pressure is applied, and we only consider the steady state regime in this study. **b,** Streaming current per channel $I_{str}/N$ as a function of the pressure gradient $\Delta P/L$ for channels in **a**, and with varying KCl concentration $c$. For each $c$, the line corresponds to the best linear fit. **c,** Electro-osmotic mobility μ as a function of the KCl concentration (linear-logarithmic coordinates; dashed line is a guide to the eye). Error bars represent: **a** - Error in the currents measured during temporal evolution is ± 0.1 pA, **b** - standard error, **c** - uncertainty on the fit value.

We now explore in more detail the pressure-driven current $I_{str}$ without applying bias ($\Delta V = 0$). The steady-state current reached after each pressure increment is shown in Fig. 2a. For concentrations between 1 and 300 mM, $I_{str}$ is found to increase linearly with the driving force, i.e., pressure gradient (Fig. 2b). From the measured slopes we calculate the streaming (electro-osmotic) mobility $\mu = I_{str}/(NA\Delta P/L)$, where $A = wh_0$ is the slit cross-sectional area. The streaming mobility displays a weak dependence on the salt concentration (Fig. 2c) and varies by less than 50% if $c$ is increased by a factor of 300. Beyond the variation, the absolute value of μ is surprisingly high, of the order of $10^{-7}$ m$^2$ V$^{-1}$ s$^{-1}$, approximately twice as large as the bulk potassium electrophoretic mobility $\mu_{K+} = 7.6 \times 10^{-8}$ m$^2$ V$^{-1}$ s$^{-1}$, here used as a reference value. This is an order of magnitude larger than the streaming mobilities reported in the literature (for example, the streaming mobility for SiO$_2$ channels is ~ 0.1 $\mu_{K+}$). Generally, the streaming mobility, which is the relevant physical quantity, is commonly translated in terms of the zeta potential, which has the dimension of an electrostatic potential. If we extend this concept to our case using the bulk water properties (viscosity η = 1 mPa·s and dielectric permittivity ε ≈ 80), the apparent Zeta potential, $\zeta = -\mu\eta/\varepsilon\varepsilon_0$, is roughly - 0.4 V, at least ten times larger than the typical values in the literature[6,24,25], which are usually of the order of $kT/e \approx 25$ mV. Moreover, recent studies of confined water indicate that its permittivity can be dramatically suppressed[21] to $\varepsilon_r \sim 2$ whereas η remains on the order of the bulk value[2]. This would translate into an apparent ζ of - 16 V! In our opinion, such a large value does not reflect anomalously high surface potential of the graphite, it merely highlights the high streaming mobility arising from the fast transport of ions by water at molecular distances from surfaces.

Our devices also allow investigation of how the pressure-induced current couples to electric forces at these molecular scales. To this end, we explore the pressure-driven streaming current as a function of applied electric field (voltage bias). Fig. 3a shows the time response of $I_{str}$ when applying pressure and, simultaneously, $\Delta V$. The results shown in Figs. 3a,b reveal a considerable coupling between the electric bias and pressure: the pressure-induced streaming current is increased by more than 100% for $\Delta V = 50$ mV as compared to the reference streaming current at zero $\Delta V$. This means that the effects of $\Delta P$ and $\Delta V$ do not simply add. Furthermore, Fig. 3b shows $I_{str}$ measured for $c = 300$ mM as a function of $\Delta P/L$ for fixed $\Delta V$ ranging from - 75 to 75 mV. The key observation here is that the current always remains proportional to the pressure gradient, independent of applied bias, but the slope - the streaming mobility $\mu(\Delta V)$ - varies with $\Delta V$, that is, $I_{str} = \mu(\Delta V) \times A \times N \times \Delta P/L$. This linear pressure response highlights that the streaming current originates from the hydrodynamic transport of ions, while its voltage dependence unveils an unexpected interplay between the mechanical and electric driving forces. The resulting mobilities, normalized by $\mu_{K^+}$, are plotted in Fig. 3d for graphite channels, as a function of $\Delta V$ and $c$.

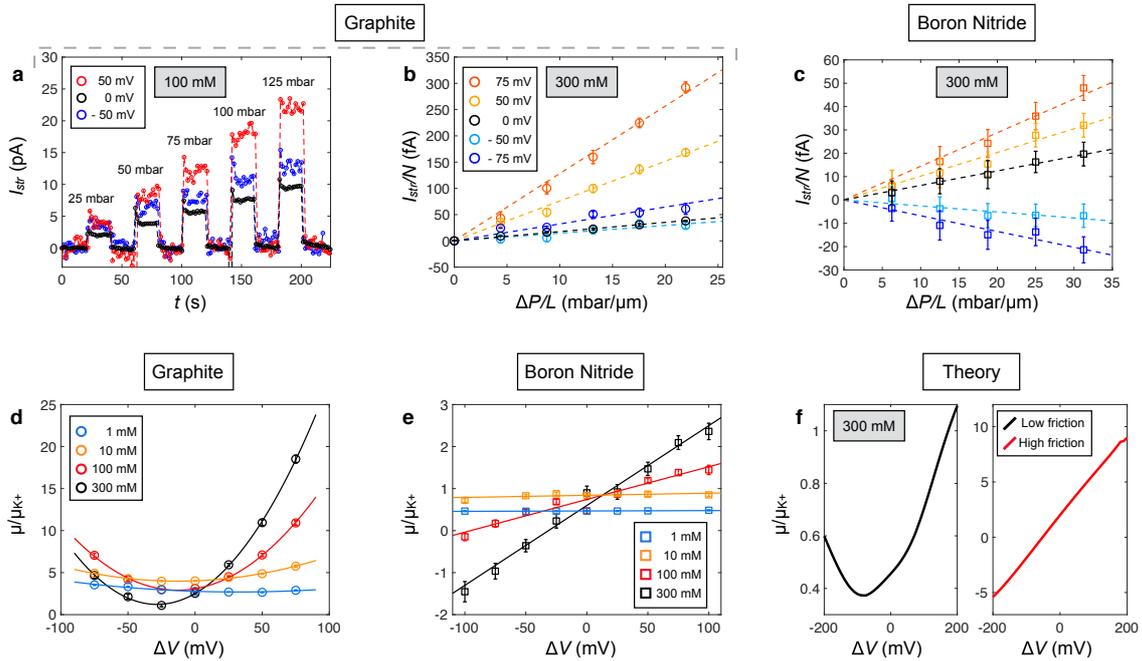

**Fig. 3 | Streaming current for different biases and channel materials. a**, Pressure-driven $I_{str}$ for a bilayer graphite device at different $\Delta V$. $L = 5.7 \pm 0.1$ μm; KCl concentration of 100 mM. The pressure applied for 20 s intervals is gradually increased to 125 mbar in 25 mbar steps. **b,** Streaming current per channel for the same device as a function of $\Delta P/L$ (bias $\Delta V$ ranges from -75 to 75 mV; colour coded). **c,** Streaming current for similar devices but with BN walls; same experiments and colour coding as in **a** and **b**. **d,** Streaming mobility (normalized by the K$^+$ electrophoretic mobility) as a function of $\Delta V$ for different KCl concentration for the graphite devices. Curves are the quadratic fits. **e**, Same as **d** but with BN channels. Linear fits; $L = 16 \pm 0.1$ μm. **f**, extended PNP prediction for the streaming mobility using different water/ions/wall friction coefficients, with a factor of 100 between low and high friction. Low friction reproduces the quadratic gating observed for graphite (panel d), while high friction leads to linear gating observed for BN (panel e). Detailed parameters and geometry used in the model are given in

the Methods section 'extended Poisson-Nernst-Planck theory' and the Extended Data Figs. 6 to 10. Error bars represent: **a-c** measurement uncertainty, **d-e** uncertainty on the fit value.

To gain more information, we compare the streaming effects in graphite channels with those in similar channels but made from another material, hexagonal boron nitride (BN). It shares a similar crystal structure and atomic flatness as graphite but is electrically insulating[26]. Our BN devices were fabricated using the procedures described above for graphite and had the same parameters including $h_0$. Overall, they exhibited behaviour similar to that of graphite devices: $I_{str}$ varied linearly with $\Delta P$ (see Fig. 3c, Extended Data Figs. 2 and 4); and the slope (streaming mobility) was again tuneable by applied bias. The extracted dependence $\mu(\Delta V)$ for BN is shown in Fig. 3e. Surprisingly, the functional dependences of $\mu(\Delta V)$ dramatically differ for the two materials. For graphite, $\mu$ shows a quadratic response to electric bias (Fig. 3d) whereas for BN it is essentially linear over the entire studied range (Fig. 3e). The data can be described by:

$$\text{for graphite, } \mu(\Delta V) = \mu_0 \left[1 + \alpha \left(\frac{\Delta V - V_{min}}{V_{ref}}\right)^2\right] \quad (1)$$

$$\text{for BN, } \mu(\Delta V) = \mu_0 \left[1 + \beta \frac{\Delta V}{V_{ref}}\right] \quad (2)$$

where $V_{ref} = kT/e \approx 25$ mV is the thermal voltage, $\mu_0$ is a mobility and $\alpha$ and $\beta$ are dimensionless parameters accounting for the voltage response. Typically, $V_{min}$ is found of the order of $V_{ref}$ and decreases with $c$; the voltage susceptibility $\alpha$ increases linearly with concentration (Extended Data Figs. 5a,b), reaching a value close to unity for high $c = 300$ mM. The characteristic mobility $\mu_0$ is typically of order of $\mu_{K+}$ for both systems. However, similar to $\alpha$ in graphite, the bias susceptibility $\beta$ for BN increases linearly with $c$ (Extended Data Fig. 5c). Due to the linear voltage coupling, the sign of the streaming current for BN can be inverted for negative biases (Extended Data Fig. 4). For both materials the sensitivity of $I_{str}$ to voltage bias is very large, in contrast to any other known control or gating mechanism[27–31]. For graphite channels, a relatively small voltage ($\Delta V \approx 75$ mV) yields streaming mobilities which are up to ~ 20 times larger than the bulk potassium mobility, taken as a reference. This corresponds to zeta potentials up to 2 V assuming the bulk water properties, and ~ 100 V if using the confined-water permittivity $\varepsilon_r$. Although the effect is still large for BN, it is substantially smaller compared to that in graphite channels. This observation echoes the smaller slip length for water on BN as compared to graphite[8,23,32].

Altogether our findings suggest that the applied bias acts as a gate for pressure-driven streaming currents. To rationalize these results, it is *a priori* tempting to understand the behaviour in terms of capacitive gating, as assumed *e.g.* for flowFET type devices[27]. Such approaches are, however, not able to capture our experimental observations, notably the contrasting voltage dependence of the gating for graphite and BN. Furthermore, it neglects the electro-hydrodynamic couplings at play for the ion and water transport under this ångström-scale confinement, which are usually described in terms of the Poisson-Nernst-Planck-Stokes (PNPS) framework. This approach is commonly used to describe ionic transport in biological or artificial channels; however, this model is unable to account for our overall observations. In particular, although it can yield to some extent a bias dependence of µ, it cannot account for the qualitatively different behaviour of graphite versus boron nitride, as summarized in Eqs. (1)-(2). This indicates that some key ingredients are missing in the PNPS model. In such a strong confinement, Stokes equation becomes irrelevant to describe the flow within the water/ions layer. In particular, the confinement leads to strong and direct interactions of the moving ions and water molecules with the walls and we account for this effect by simply considering frictions between water, ions and the walls. This translates into an effective water-wall friction depending on the ion concentrations, which may be described as $\lambda_w(\rho_+,\rho_-) = \lambda_0 + h_0(\kappa_+\rho_+ + \kappa_-\rho_-)$, where $\lambda_0$ is the bare (ion-free) friction coefficient for water, $\kappa_\pm$ are coefficients characterizing the ions' contribution to the friction, and $\rho_\pm$ are the ion concentrations. The full model for our channel geometry is detailed in the Methods section 'extended Poisson-Nernst-Planck theory' and Extended Data Figs. 6 to 10. This extended PNP model is found to reproduce qualitatively most of the experimental observations. First, it leads to streaming currents that are linear in $\Delta P$ (Extended Data Fig. 7). Second, the model can reproduce a large increase in the streaming mobility under applied bias, Fig. 3f. Note that this is in contrast to the standard consequences of concentration polarization, which tend to counteract the effects. Even more strikingly, as highlighted in Fig. 3f, the model yields different functional dependences $\mu(\Delta V)$ that depend on the friction of both water and ions on the different materials: a low water/wall and ion/wall friction leads to a quadratic gating of the streaming mobility, as observed for graphite, whereas large frictions lead to a practically linear behaviour, as indeed observed for BN. The theory indicates that this behaviour results from a subtle and intertwined effect of concentration dependent flow and ion concentration profile across the channels in the presence of voltage. Accordingly, the different material response observed in Fig. 3 for channels made from BN and graphite can be traced back to the difference of molecular friction of water and ions on these two materials. This is in agreement with expectations for the friction of water on these two materials[8,23,32]. A remarkable feature of this framework is that the observed non-linear bias response thus takes its roots in the fundamental nature of interactions between confining walls, water molecules and ions. For instance, the minimum mobility seen in Fig. 3d arises from the slight asymmetry in device geometry which is at the origin of different frictions (induced by the confinement) and modifies locally the transport rates of ions on each side. Although the theory reproduced qualitatively our experimental results (Extended Data Figs. 8 to 10), it is still not able to account for the large amplitude of bias effect on graphite and further work is necessary to reach better agreement. In particular, the effects of strong confinement (including the suppressed dielectric constant) are expected to modify ions' adsorption[33] as well as water and ion dynamics; furthermore the metallicity of graphite can modify substantially the ionic interactions and ultimately modify their concentration. *Ab initio* molecular simulations such as those by Tocci *et al.*[23] could provide further insight into the effect of ions on water friction, beyond the simple picture proposed here.

Our experimental system allows us, for the first time, to probe a purely two-dimensional water/ions flow, a configuration very different from the one-dimensional transport of nanotubes. Thanks to the lateral extension of the ångström-channels, streaming currents under molecular confinement become measurable. Hence, such devices are an interesting platform to mimic the behaviour of biological channels in terms of stimuli responsive behaviour such as voltage gating, where ions are driven through ångström-scale confinement by coupled osmotic pressure and electric forcing. This is of particular relevance for gaining new insights into the electro-mechanical coupling at the root of the mechanosensitivity observed in recently discovered biological ionic channels[4] (TRAAK, TREK, PIEZO). Furthermore, the observed friction-based electric gating opens a new route to achieve flow-control under extreme confinement where small voltages induce strong responses, which would constitute an important step towards building nanofluidic circuits responding to external stimuli.

*Acknowledgments*

T.M. and L.B. acknowledge funding from ANR project Neptune. B.R. acknowledges Royal Society Fellowship, a L'Oréal Fellowship for Women in Science, and EPSRC grant EP/R013063/1. A.S. acknowledges funding from the European Union's H2020 Framework Programme/ERC Starting Grant agreement number 637748 - NanoSOFT. L.B. acknowledges funding from the European Union's H2020 Framework Programme/ERC Advanced Grant agreement number 785911 - Shadoks. A.R.P. acknowledges funding from the European Union's Horizon 2020 Framework Program/European Training Program 674979 - NanoTRANS. S.A.D. was funded by a scholarship from the University of Engineering and Technology, Lahore Pakistan. A.K., B.R. and A.K.G. were supported by Lloyd's Register Foundation and European Research Council (ARTIMATTER). T.M. thanks S. Blin and H. Yoshida for assistance.

*Author contribution*

B.R., L.B., and A.S. designed and directed the project. A.K., B.R. and S.A.D. fabricated the devices. T.M. performed the measurements and their analysis. A.R.P., T.M. and L.B. provided theoretical support. T.M., L.B., B.R., A.K., A.R.P. wrote the manuscript with the inputs from A.K.G. All authors contributed to discussions.

*Author information*

The authors declare no competing interests.

*Methods*

**Device fabrication**

Our devices were fabricated following the previously reported procedures[2,3]. In brief, a free-standing silicon nitride (SiN) membrane of around 500 nm in thickness provided mechanical support and also served to separate the two reservoirs connected by the channels. On the membrane, a rectangular hole of ~ 3 μm x 26 μm was defined by lithography and plasma etching. The channels were made by van der Waals assembly of three layers - bottom, spacer, top - of 2D crystals such as graphite or hexagonal boron nitride (BN). First, a bottom layer of around 10 to 50 nm thick graphite or BN was transferred onto the hole in the SiN membrane and etched from the back side, which projected the hole into the bottom layer. Following this, pre-patterned bilayer graphene spacers (~ 6.8 Å thick) in the form of parallel ribbons of ~ 130 nm wide and separated also by ~ 130 nm were transferred onto the bottom crystal and aligned perpendicular to the long axis of the rectangular hole. Finally, a thick (~ 100 to 150 nm) top crystal of graphite or BN was transferred onto the spacers covering the hole (Extended Data Fig. 1). The top crystal defined the length of the channels that formed on both sides of the hole.

**Streaming current measurements**

Extended Data Figs. 3a-c show the streaming current measurements as a function of the applied pressure for a sample containing no channels (varying from 0 to 250 mbar). The pressure is applied *via* an Elveflow - AF1 pressure pump (we denote a positive $\Delta P$ for a pressure applied through the hole on SiN) and $\Delta V$ is controlled via a patch-clamp amplifier (Molecular Device - Axopatch 200B) with the ground electrode on the top side. For a sample containing no channels, we did not detect any significant current. Extended Data Fig. 3d-f compares the streaming current measured for the control sample and a graphite device containing 200 channels. In the case of graphite channels, the streaming current is four orders of magnitude larger than the noise measured in the control sample.

In order to investigate the pressure dependence of the streaming current, we performed the streaming current measurements applying the pressure successively on each sides of the membrane. The inversion of the pressure gradient fully reverts the streaming current sign as presented in the Extended Data Fig. 2; this confirms the linear dependence of the streaming current on the mechanical forcing.

The molecular streaming current $I_{str}$ as a function of the pressure gradient $\Delta P/L$ is shown in Extended Data Fig. 4 for both graphite and BN devices and for different KCl concentrations and applied voltages. The streaming current varies linearly with the driving force $\Delta P/L$.

**Extended Poisson-Nernst-Planck theory**

**1. Governing equations**

At scales greater than ~1 nm, the influence of water motion on the ionic fluxes is accounted for by (1) appending Stokes' equation for the solvent velocity to the typical Poisson-Nernst-Planck (PNP) description of the ionic transport and (2) including an ionic drift velocity set by the balance of forces between the electric force on the individual ion and the frictional force between the ion and water in the Nernst-Planck parameterisation of the solute fluxes. Both assumptions are inapplicable here due to the extreme confinement scale of the channels considered, which approaches the diameter of the water molecules and hydrated ions themselves. In particular, application of the Stokes equation to predict the hydrodynamic velocity relies on the assumption of a spatially homogeneous and isotropic scalar viscosity, an assumption that cannot be valid when a single layer of water molecules is present. *A priori*, we would expect strong interaction between the ions and walls and the water molecules and walls. The former supposition is supported by the results of Esfandiar *et al.*[3], where the chloride mobility in both graphite and BN devices of the type examined here was observed to be reduced by approximately 65% compared to bulk. The latter is supported by the present results when combined with the simple, first-principles model detailed below.

As noted above, the traditional ionic and hydrodynamic force balances, leading to the typical parameterisation of the drift velocity and the Stokes equation, respectively, can no longer be sufficient to describe the coupled ion-water transport in one-to-two layers confinement owing to the substantial interaction with the confining material. As a simple, first-principles approach, we consider the force balances on the individual ions and on a control volume of infinitesimal length along the slit containing both ions and water molecules. We include three phenomenological forces, the frictional interactions of (1) water with walls, (2) ions with walls and (3) ions with water. We emphasize that this is the simplest possible coherent approach to capture the modification in the qualitative behaviour of the ion dynamics owing to the extreme confinement. Quantifying the friction to achieve a more quantitatively accurate treatment would likely necessitate more in-depth modelling (*e.g. ab initio* molecular dynamics).

Including the ion-wall interaction, a force balance on an individual ion gives:

$$0 = \pm e(-\partial_x \phi) - \xi_\pm (v_\pm - v_w) - \lambda_\pm v_\pm, \tag{S1}$$

where $v_\pm$ is the velocity of the positive or negative ion species, $v_w$ is the water velocity and $\phi$ is the electrostatic potential. From left to right, the terms represent (1) the electric body force on the positive or negative ion, (2) the frictional force of the water on the ion, parameterised by friction coefficients $\xi_\pm$ for the cation and anion species and (3) the frictional force of the wall on the ions, parameterised by friction coefficients $\lambda_\pm$. Note that we have assumed that all of the ions interact appreciably with the walls, a reasonable assumption here given the extreme confinement. We solve for the ion velocity $v_\pm$ to obtain:

$$v_\pm = \pm \mu_\pm (-\partial_x \phi) + \alpha_\pm v_w. \tag{S2}$$

We have introduced the ionic mobilities $\mu_\pm \equiv e/(\xi_\pm + \lambda_\pm)$ and the normalized water-ion friction coefficients $\alpha_\pm \equiv \xi_\pm/(\xi_\pm + \lambda_\pm) \in (0,1)$. The former parameters are constrained by the experimental results of Esfandiar *et al.*[3]; the latter parameters characterise how effectively the drag of the water flow is able to overcome frictional resistance on the ions from the wall and engender ionic transport. We note that the definition of $\alpha_\pm$ may be rearranged to give $\lambda_\pm/\xi_\pm = (1-\alpha_\pm)/\alpha_\pm$. This indicates that a value $\alpha_\pm \ll 1$ corresponds to stronger ion-wall than ion-water friction, while values of $\alpha_\pm \sim 1$ indicates relatively weaker ion-wall than ion-water interaction.

From the above definitions, we see that the sums of the ion-water and ion-wall friction coefficients are constrained by the experimentally measured mobilities reported in Esfandiar et al.[3], $\xi_\pm + \lambda_\pm = e\mu_\pm^{-1}$, while the relative importance of the ion-wall and ion-water interactions, characterized by the ratios $\lambda_\pm/\xi_\pm = (1-\alpha_\pm)/\alpha_\pm$, are not.

We next consider the force balance on a control volume (CV) of width and height equal to the channel width $w$ and height $h_0$, respectively, and of infinitesimal length $\delta x$ in the along-slit direction. The total volume of the CV is then $\delta V \equiv wh\delta x$. The total electric body force is given by $e(\rho_+ - \rho_-) \times (-\partial_x\phi) \times \delta V$, and the net pressure force is given by $wh \times (-\partial_x P)\delta x$. In the preceding, $\rho_\pm$ are the ionic densities (per unit volume) at the position $x$ coincident with the centre of the CV (so that in the reservoirs $\rho_\pm = \mathcal{N}_a c$ with $\mathcal{N}_a$ the Avogadro number), and $P$ is the pressure. The total frictional force due to ion-wall interactions is $-(\rho_+\lambda_+ v_+ + \rho_-\lambda_- v_-) \times \delta V$. Finally, we introduce a coefficient $\lambda_0$ characterizing the frictional interaction of water molecules with the walls such that $-\lambda_0 v_w$ is the force per unit wall area acting on the water molecules, and $-\lambda_0 v_w \times w\delta x$ is the total frictional force on the CV due to water-wall interaction. The force balance on the CV thus gives:

$$0 = e(\rho_+ - \rho_-)(-\partial_x\phi)\delta V + (-\partial_x P)\delta V - (\rho_+\lambda_+ v_+ + \rho_-\lambda_- v_-)\delta V - \frac{\lambda_0}{h}v_w\delta V. \quad (S3)$$

Before solving the above for the water velocity $v_w$, we use Eq. S2 and the definitions of $\mu_\pm$ and $\alpha_\pm$ to rewrite the total ion-wall friction force per unit volume $\delta V$, $\rho_+\lambda_+ v_+ + \rho_-\lambda_- v_-$, as:

$$e(\rho_+ - \rho_-)(-\partial_x\phi) - e(\alpha_+\rho_+ - \alpha_-\rho_-)(-\partial_x\phi) + (\kappa_+\rho_+ + \kappa_-\rho_-)v_w, \quad (S4)$$

where we have defined $\kappa_\pm \equiv e\alpha_\pm(1-\alpha_\pm)/\mu_\pm$ and made use of the identities $\lambda_\pm\mu_\pm \equiv e(1-\alpha_\pm)$ and $\lambda_\pm\alpha_\pm \equiv \kappa_\pm$. We insert this result into Eq. S3 and solve for $v_w$ to obtain:

$$v_w = K_{app}(\rho_+, \rho_-)[(-\partial_x P) + e(\alpha_+\rho_+ - \alpha_-\rho_-)(-\partial_x \phi)], \qquad (S5)$$

where $K_{app}(\rho_+, \rho_-)$ is a concentration-dependent apparent hydraulic permeance, given by:

$$K_{app}(\rho_+, \rho_-) \equiv \frac{1}{\frac{\lambda_0}{h} + \kappa_+\rho_+ + \kappa_-\rho_-}. \qquad (S6)$$

In order to better interpret the significance of the parameter $\alpha_\pm$ and the non-intuitive form in which the electric field appears in Eq. S5, we use the above results to calculate the difference of the electric force $f_e^\pm$ and the ion-wall friction force $f_{ion-wall}^\pm$ on a given ionic species:

$$f_e^\pm - f_{ion-wall}^\pm = \pm e\alpha_\pm\rho_\pm(-\partial_x\phi) - \kappa_\pm\rho_\pm v_w. \qquad (S7)$$

Let us discuss two extreme limits. When $\alpha_\pm = 0$, $\xi_\pm/\lambda_\pm = 0$, indicating that only ion-wall (rather than ion-water) friction is relevant. Further, from the above definition, $\kappa_\pm \propto \alpha_\pm(1 - \alpha_\pm) = 0$, and the net (electric less ion-wall friction) force vanishes. Thus, in this case, all of the electric force on the given ionic species in the CV is balanced by the strong ion-wall interaction such that the given ionic species does not communicate any electric force to the water molecules. (See Eq. S5 with $\alpha_+$ and/or $\alpha_-$ set to zero.)

On the other hand, when $\alpha_\pm = 1$, $\lambda_\pm/\xi_\pm = 0$, indicating only ion-water friction is relevant, and all of the electric force on the ions is communicated to the water molecules such that $f_e^\pm - f_{ion-wall}^\pm = \pm e\rho_\pm(-\partial_x\phi)$. (Again, $\kappa_\pm \propto \alpha_\pm(1 - \alpha_\pm) = 0$.)

We emphasize that the behaviour described in Eqs. S5 and S6 is in strong contrast to what is observed for conduits with confinement scale (radius or height) > ~1 nm, in which Hagen-Poiseuille holds[19]. In this case, we would have a concentration-independent permeance $K_{HP} = h_0/\lambda_0$ and a net electric driving force equal to the total electric driving force $e(\rho_+ - \rho_-)(-\partial_x\phi)$. $K_{HP}$ is recovered in the high water friction limit, $\lambda_0/h \gg \kappa_+\rho_+ + \kappa_-\rho_-$, and both $K_{HP}$ and the total electric driving force are recovered outside of confinement where $\alpha_\pm = 1$ (equivalent to no ion-wall friction: $\lambda_\pm = 0$).

It is necessary to use Eq. S5, instead of Hagen-Poiseuille, in order to capture the full range of qualitative behaviour observed in the experimental $\mu(\Delta V)$ curves. This emphasizes the importance of the two-dimensionality of the flow, resulting in a strong frictional interaction between the channel walls, water and ions.

We insert Eq. S5 into the general Nernst-Planck parameterisation for the ionic fluxes, $j_\pm = D_\pm(-\partial_x \rho_\pm) + v_\pm \rho_\pm$, to obtain:

$$j_\pm = \mu_\pm \left[\frac{k_B T}{e}(-\partial_x \rho_\pm) \pm \rho_\pm(-\partial_x \phi)\right] + \alpha_\pm v_w \rho_\pm, \tag{S8}$$

where we have made use of the Einstein relation, $D_\pm = k_B T \mu_\pm / e$.

At steady-state, the conservation equations become:

$$\frac{d(hv_w)}{dx} = 0, \frac{d(hj_\pm)}{dx} = 0. \tag{S9}$$

Finally, the electrostatic potential $\phi$ is related to the total charge density $e(\rho_+ - \rho_-)$ via the Poisson equation:

$$\partial_x[\epsilon\epsilon_0 h(-\partial_x \phi)] = he(\rho_+ - \rho_-). \tag{S10}$$

## 2. Model geometry and boundary conditions

As we are mainly interested in capturing the qualitative features of the ionic current response, we adopt a simplified one-dimensional geometry. The model geometry adopted here is sketched in Extended Data Fig. 6. A slit of uniform height $h_0 = 7$ Å and length $L = 5$ μm connects two reservoirs of divergent geometry. It is necessary to include the reservoirs in some capacity in our calculations in order to capture the entrance/exit effects associated with the discontinuous change in ionic mobility as the ions enter/exit the channel. The rate of divergence of the reservoir heights is asymmetric, qualitatively mimicking the asymmetry of the experimental geometry. The height profile $h(x)$ is given by:

$$\frac{h(x)}{h_0} = \begin{cases} \Gamma_\ell\left[\left(\frac{x}{L}\right)^2 + \frac{1}{2}\right], & x \in \left(-\infty, -\frac{L}{2}\right) \\ 1, & x \in \left[-\frac{L}{2}, +\frac{L}{2}\right] \\ \Gamma_r\left[\left(\frac{x}{L}\right)^2 - \frac{1}{2}\right], & x \in \left(+\frac{L}{2}, +\infty\right). \end{cases} \quad (S11)$$

$\Gamma$ is the rate of divergence of the confinement: the larger $\Gamma$ is, the more abrupt is the transition to the open reservoir. We take $\Gamma_l = 5$ and $\Gamma_r = 20$. While the magnitudes of $\Gamma_l$ and $\Gamma_r$ influence the quantitative predictions of the model, the qualitative behaviour of the mobilities are similar so long as $\Gamma_l < \Gamma_r$.

We impose the reservoir conditions at $x = \pm\infty$. In the left reservoir, we apply a voltage and pressure:

$$\phi(x = -\infty) = \Delta V, \quad (S12)$$
$$P(x = -\infty) = \Delta P. \quad (S13)$$

In the right reservoir, the voltage and pressures are held fixed at reference values arbitrarily set to zero:

$$\phi(x = +\infty) = 0, P(x = +\infty) = 0. \quad (S14)$$

The total ionic density in both reservoirs is held fixed at $\rho_{res} = 2\mathcal{N}_a c$, and both reservoirs are assumed to be electroneutral, such that:

$$\rho_\pm(x = \pm\infty) = \frac{\rho_{res}}{2}. \quad (S15)$$

### 3. Variation of ion mobilities $\mu_\pm$ and normalized water-ion friction coefficients $\alpha_\pm$

We impose the following profiles for the ionic mobilities:

$$\mu_\pm = \left(\mu_\pm^{bulk} - \mu_\pm^{conf}\right)\left[1 - \frac{\tanh\left(\frac{x+\frac{L}{2}}{\lambda_{adj}}\right) - \tanh\left(\frac{x-\frac{L}{2}}{\lambda_{adj}}\right)}{2}\right] + \mu_\pm^{conf}, \quad (S16)$$

with an adjustment length $\lambda_{adj} = 0.3$ nm. In order to qualitatively account for the reduction in chloride mobility, we take $\mu_-^{conf} = 0.5\mu_-^{bulk}$. Similarly, we impose for the normalized water-ion friction coefficients:

$$\alpha_\pm = \left(1 - \alpha_\pm^{conf}\right)\left[1 - \frac{\tanh\left(\frac{x+\frac{L}{2}}{\lambda_{adj}}\right) - \tanh\left(\frac{x-\frac{L}{2}}{\lambda_{adj}}\right)}{2}\right] + \alpha_\pm^{conf}. \tag{S17}$$

**4. Results**

Calculations were performed using the finite element method (COMSOL). Extended Data Fig. 7 shows the results of the above model for low water-wall ($\lambda_0/h_0 = 10^{11}$ kg m$^{-3}$ s$^{-1}$) and water-ion ($\alpha_+ = 1 \Leftrightarrow \lambda_+/\xi_+ = 0$; $\alpha_- = 0.7 \Leftrightarrow \lambda_-/\xi_- \approx 0.43$) frictions in panels a through c, and those for high water-wall ($\lambda_0/h_0 = 10^{13}$ kg m$^{-3}$ s$^{-1}$) and water-ion ($\alpha_+ = \alpha_- = 0.01 \Leftrightarrow \lambda_+/\xi_+ = \lambda_-/\xi_- = 99$) frictions in panels d through f. We first note that in both cases we reproduce the linear dependence of the streaming current on the pressure gradient for both zero and nonzero applied voltages (Extended Data Figs. 7b and e), in agreement with experiments (Figs. 3b and c, main text).

The low friction results produce a quadratic dependence of the streaming mobility on the applied voltage with a minimum mobility occurring for $\Delta V = V_{min} < 0$ (Extended Data Figs. 7c). This qualitative behaviour is in agreement with the experimental results obtained for graphite (Fig. 3d, main text). Likewise, the high friction results reproduce the linear dependence of the streaming mobility on $\Delta V$ (Extended Data Fig. 7f) that is observed experimentally in BN (Fig. 3e, main text). The frictional characteristics of these results are consistent with the typically much lower friction (larger slip lengths) observed on graphite than in BN[7,23,32]. We note that, in addition to taking low to moderate values of $\lambda_\pm/\xi_\pm \sim 0 - 1$, it is necessary to take $\alpha_+ > \alpha_-$ in order to recover the qualitative behaviour of graphite. On the other hand, it is necessary to take $\alpha_+ \approx \alpha_- < \sim 0.1$ in order to recover the qualitative behaviour of BN. This suggests that frictional interaction of the wall with the ions is weaker generally in graphite, and that it is stronger for chloride than potassium. In BN, on the other hand, our results suggest that the frictional interaction of the wall with the ions is quite strong for both species.

The numerical results presented here for the low friction (graphite-like) configuration indicate that $\mu(\Delta V = 0)$ is independent of concentration, roughly consistent with the minimal variation observed in the experiments (Fig. 2c, main text). However, the linear dependence of the mobility on concentration for nonzero applied voltages (Fig. 3d-e, main text) is not observed in the model (Extended Data Fig. 7c). Conversely, at higher friction (Extended Data Fig. 7d through f), $\mu(\Delta V = 0)$ varies strongly with the concentration, as well as the gated mobility (Extended Data Fig. 7f). This suggests the possibility that the concentration, applied voltage, and friction are coupled in ways not accounted for in our simple model.

The numerical results depend crucially on the difference in water flow characteristics between the two materials through the concentration-dependent permeance given in Eq. S6. However, the nature of this dependence is highly intricate. Our numerical results indicate that, in addition to the advective current engendered by the applied pressure, the streaming current characteristics depend crucially on the modification of the electrophoretic current, $I_{ep} \propto \rho E$ (figure not shown) via the modification of the concentration and electrostatic fields by coupled voltage and pressure effects. An example of the influence of voltage on the evolution of the concentration fields in the presence of a fixed applied pressure gradient $\Delta P/L = 30$ mbar/μm and a reservoir concentration $c = 300$ mM is shown in Extended Data Fig. 8c and d. We see that both the applied pressure and voltage induce modification of the concentration profile across the channel (as well as charge separation, not shown). The modification of the concentration profile due to pressure is much stronger in BN (Extended Data Figs. 8a and c), and it is also much more sensitive to applied pressure and voltage in BN than in graphite. This latter characteristic is consistent with the smaller streaming mobilities observed in the graphite-like configuration observed in our numerical results (Extended Data Figs. 7c and f). Additionally, we see that the evolution of the concentrations under coupled $\Delta P$-$\Delta V$ forcing is quite different in the two materials; it is this difference, and the corresponding difference in the response of the advective and electrophoretic currents, that determines the difference between the two material behaviours.

There are several aspects of the observations in graphite that we are not able to reproduce: (1) the non-monotonicity of the dependence of $\mu(\Delta V = 0)$ on concentration, (2) the linear dependence of the mobility on concentration when a voltage is applied, and (3) the magnitude of the mobilities measured at high concentration under an applied voltage. Indeed, the model consistently predicts mobilities in the quadratic (graphite-like) regime that are smaller than those observed in the linear (BN-like) regime (Extended Data Figs. 7c and f). This is not an issue of the voltage range examined, as the mobilities are found to saturate or even reduce at much higher voltages. Likewise, there is much we have not included in our model, in particular, steric effects and ionic correlations generally, as well as the 'granular' nature of water, which might be important at this length scale. Nonetheless, the model does reproduce much of the key qualitative behaviour, and its success depends on the strong differences in the frictional characteristics of BN and graphite, and further on the incorporation of the retarding influence of the ions on the water transport, an effect that is exclusively two-dimensional. Thus, these results illustrate the truly two-dimensional character of the flow and the limit of the continuum description of matter.

## 5. Geometric Sensitivity

The effect of the reservoir geometry on the numerical model predictions is illustrated in Extended Data Fig. 9. In this plot we show the influence of both the relative and absolute magnitudes of $\Gamma_l$ and $\Gamma_r$ on the predicted $\mu(\Delta V)$ responses for both the low friction (graphite-like) and high friction (BN-like) configurations. Between the blue and yellow curves, we vary the absolute magnitudes of $\Gamma_l$ and $\Gamma_r$ by an order-of-magnitude while keeping the ratio $\Gamma_l/\Gamma_r$ fixed. We see that the magnitudes of $\Gamma_l$ and $\Gamma_r$ have no influence on the qualitative (linear or quadratic) behaviour of the mobility curves, and have only a slight quantitative influence in the graphite configuration. We also vary the ratio $\Gamma_l/\Gamma_r$ (red and purple versus blue and yellow curves). In the graphite response, we see that the minima in the red and purple ($\Gamma_l/\Gamma_r = 1/8$) and the blue and yellow ($\Gamma_l/\Gamma_r = 1/4$) curves are coincident, even as we vary the absolute magnitudes of $\Gamma_l$ and $\Gamma_r$ by an order-of-magnitude. This indicates that in our model, for fixed values of the friction coefficients, the asymmetry determines the location of the minimum mobility in graphite. Likewise, in the BN curves, we see that the asymmetry is the only geometric characteristic that determines the slope of the $\mu(\Delta V)$ curve.

As a final note on the model geometry, a one-dimensional model of the type we have applied here is strictly valid only if the slope verifies $|\partial_x h| \ll 1$. Formally, this condition is not satisfied deep in the reservoirs. However, variations of the various profiles in the reservoir occur over length scales which are found to be at most of order the channel length L, so that $|\partial_x h| < \Gamma h_0/L$, which remains very small. Note furthermore that reservoirs are included merely to qualitatively capture the influence of (1) the device asymmetry, and (2) the entrance/exit effects associated with the abrupt change in anion mobility at the entrance and exit of the ångströslit. Previous work using this approach to include the reservoirs within a one-dimensional PNPS model was successful in capturing the nontrivial qualitative behaviour of the ionic current under applied pressures and voltages[34].

**6. Transition behaviour**

In Extended Data Figure 10, we show the influence of the friction parameters for high, low and intermediate friction on the gated mobilities (panels a through c) as well as the relative pressure dependence of the normalised potential $e\Delta\phi/kT$ along the channel axis (panels d through f). $\Delta\phi$ is defined as the potential variation with an applied pressure $\Delta\phi = \phi(\Delta V, \Delta P = 30 \text{ mbar/}\mu\text{m}) - \phi(\Delta V, \Delta P = 0)$. The modification of the electrostatic potential, and hence the electric field, under coupled pressure-voltage forcing contributes -along with the modification of the concentration field (Extended Data Figure 8)- to the modification of the electrophoretic current under an applied pressure. Extended Data Figures 8 and 10 illustrate the complex interplay of competing interactions that contribute to the surprisingly simple linear streaming response observed in the model.

*Data availability*

The data that support the plots within this paper and other findings of this study are available in the main text and Extended Data Figures. Additional information is available from the authors upon reasonable request.

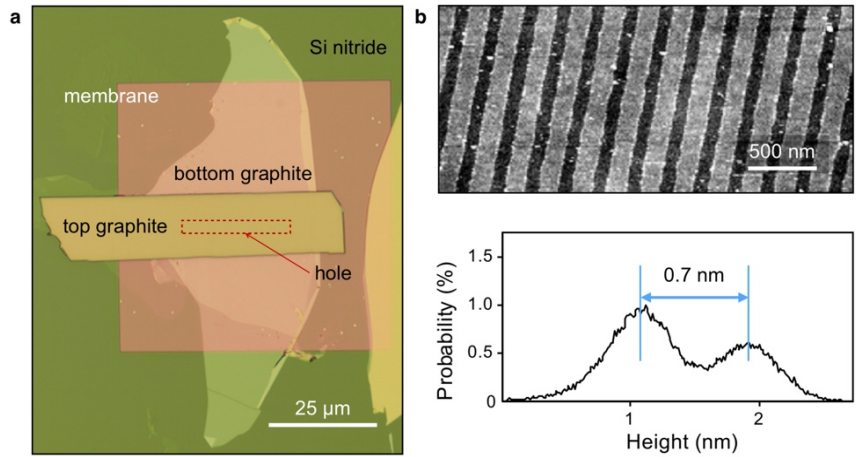

**Extended Data Fig. 1 | Ångström scale channel devices. a,** Optical image of a device with ångström channels. The square in light pink colour is silicon nitride membrane which has a rectangular hole shown in red dotted line. On top of the hole, bottom graphite, spacer and top graphite are placed. Bottom and top graphite are visible in the image in light and bright yellow colours. **b,** Atomic force microscopy image of the bilayer graphene spacer lines on the device. The histogram of the heights shows that the spacer is ~ 0.7 ± 0.1 nm thick.

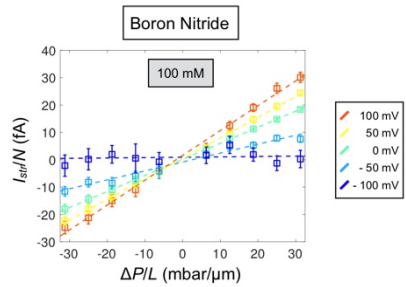

**Extended Data Fig. 2 | Gated pressure-driven current.** Streaming current per channel plotted as a function of $\Delta P/L$ with $\Delta V$ ranging between - 100 and 100 mV (colour coded from blue to red with increasing voltage difference), KCl concentration of 100 mM and with BN channels of length $L = 16 \pm 0.1$ μm.

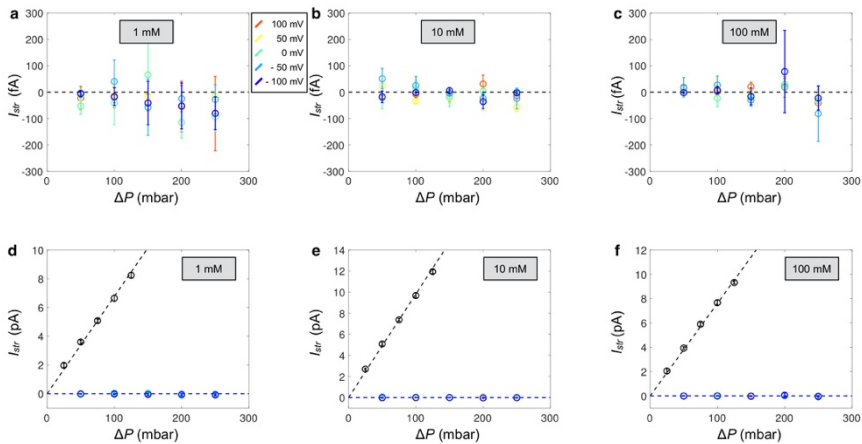

**Extended Data Fig. 3 | Control sample test. a to c,** Streaming current measured in a control sample without any channels as a function of the pressure. We varied the applied voltage from - 100 to 100 mV (colour coded from

blue to red). **d to f,** Same measurements as for **a-c** but compared with the streaming current measured with 200 graphite channels. The streaming current is around 4 orders of magnitude larger which confirms that channels remain mechanically stable, and are not delaminated under pressure.

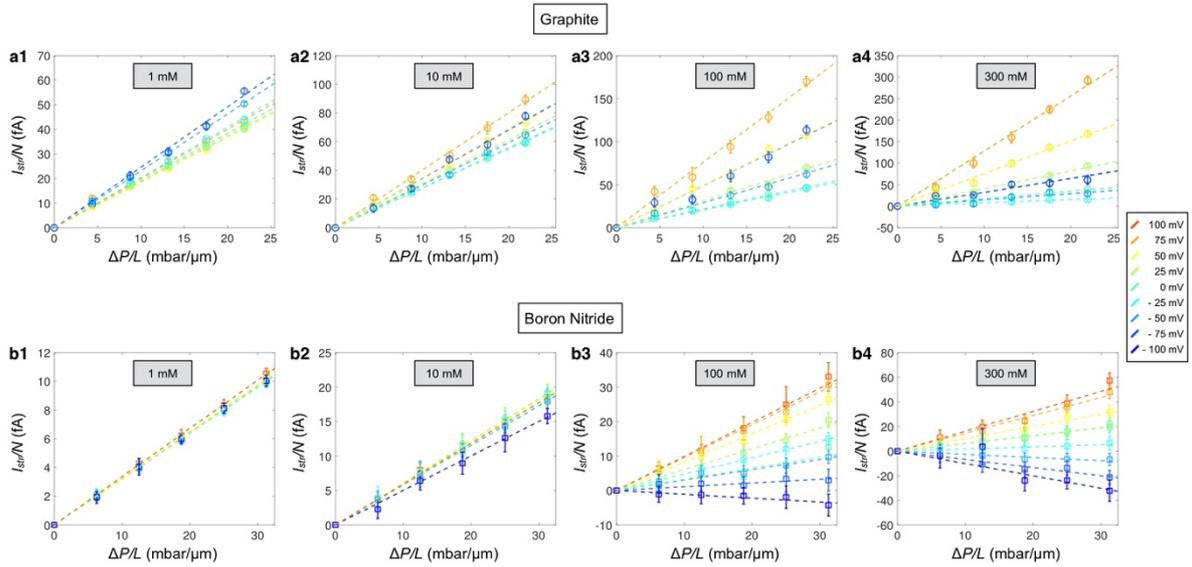

**Extended Data Fig. 4 | Gated pressure-driven current and material dependency.** Streaming current per channel plotted as a function of $\Delta P/L$ for a KCl concentration varying from 1 to 300 mM and with $\Delta V$ ranging between -100 and 100 mV (colour coded from blue to red with increasing voltage difference). **a-d,** The channel length $L$ for graphite is 5.7 ± 0.1 μm. **e-h,** For BN, $L$ = 16 ± 0.1 μm.

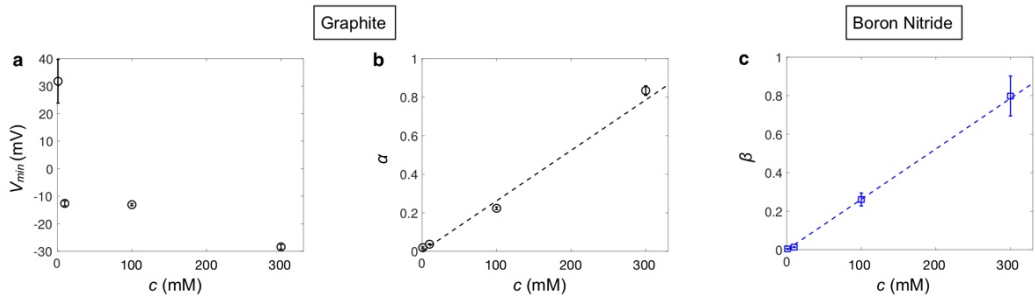

**Extended Data Fig. 5 | Concentration dependence of the fit parameters of the gate-controlled mobility.** We report the fitting parameters of the voltage gated streaming current. **a-b** correspond to the quadratic dependence of the gated streaming current observed in graphite channels (Fig. 3d, main text) and described by Eq. 1: **a,** $V_{min}$ plotted as a function of the concentration. **b,** $\alpha$ as a function of the concentration. **c,** We report the fitting parameter $\beta$ as a function of the concentration for boron nitride slits, $\beta$ describes the linear dependence of the streaming current observed for BN channels (Fig. 3e, main text) as given by Eq. 2. The dashed lines in **b** and **c** are linear fits.

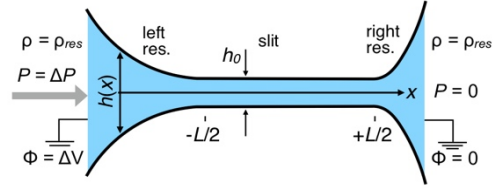

**Extended Data Fig. 6 | Geometry and effect of the asymmetry of the system.** A slit of uniform height $h_0 = 7$ Å and length $L = 5$ μm connects two asymmetric, divergent reservoirs of variable height $h(x)$. The asymmetry in the rate of divergence of the reservoir heights qualitatively mimics the asymmetry of the experimental geometry. A voltage $\phi = \Delta V$ and pressure $P = \Delta P$ are applied in the left reservoir (at $x = -\infty$); the voltage and pressure are held fixed at $\phi = 0, P = 0$ in the right reservoir ($x = +\infty$). The density in both reservoirs is held fixed at $\rho = \rho_{res}$.

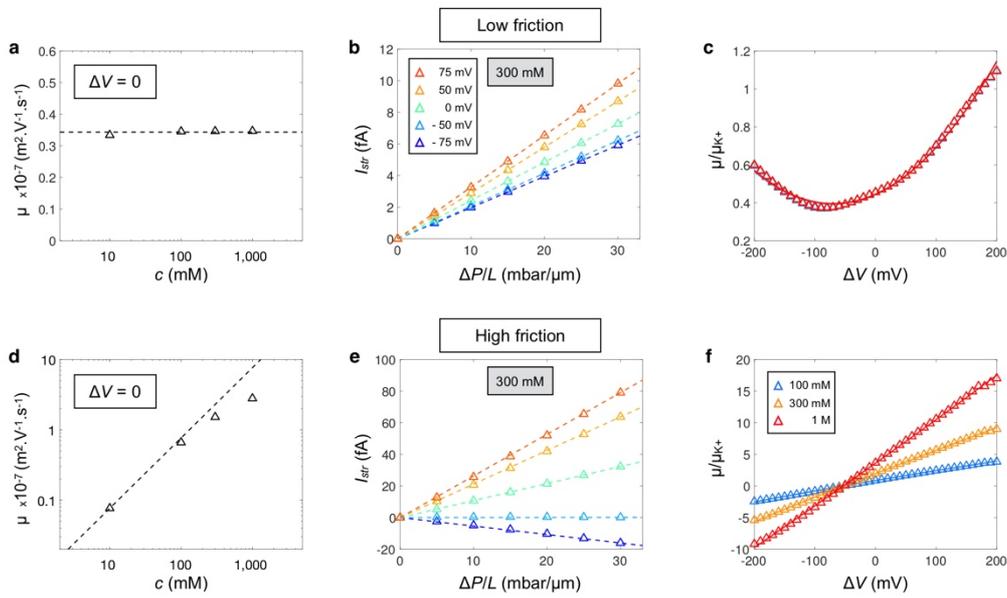

**Extended Data Fig. 7 | Prediction of the streaming current from extended Poisson-Nernst-Planck modelling. a,** Mobility without applied voltage as a function of KCl concentration in linear-logarithmic coordinates for low water-wall friction and $\alpha_+ > \alpha_-$. **b,** Streaming current per channel $I_{str}$ for 300 mM as a function of the pressure gradient $\Delta P/L$ for $\Delta V$ varying from -75 mV (blue data) to +75 mV (red data). For each voltage, the dashed line corresponds to the linear fit of the data made to extract the mobility. **c,** Streaming mobility μ normalized by the $K^+$ electrophoretic mobility $\mu_{K^+}$ and plotted as a function of the applied voltage for KCl concentration varying from 100 mM (blue data) to 1 M (red data). **d-f,** Same as in **a-c.** but with high water-wall friction and $\alpha_+ = \alpha_-$. Parameters: **a** through **c** - $\lambda_0/h_0 = 10^{11}$ kg m$^{-3}$ s$^{-1}$, $\alpha_+ = 1$, $\alpha_- = 0.7$; **d** through **f.** $\lambda_0/h_0 = 10^{13}$ kg m$^{-3}$ s$^{-1}$, $\alpha_+ = 0.01$, $\alpha_- = 0.01$. Dashed lines in **a** and **d** are guides to the eye corresponding to a constant value of μ and a linear variation with concentration, respectively

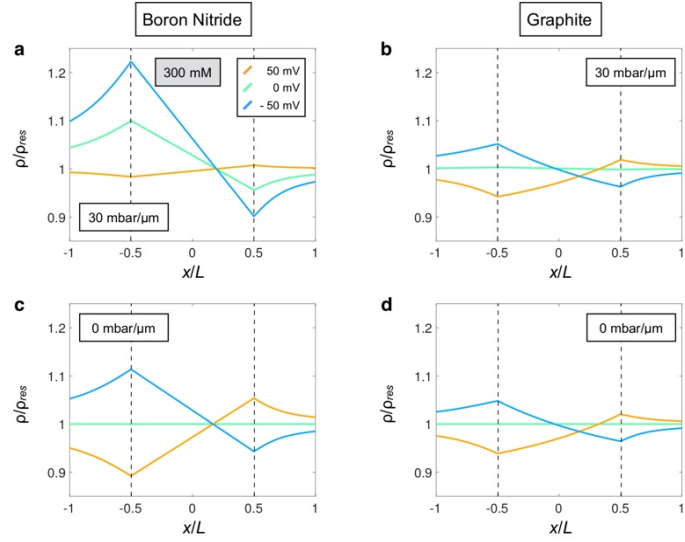

**Extended Data Fig. 8 | Total ionic concentration profiles from extended Poisson-Nernst-Planck modelling.** Total ionic concentration profiles as a function of the normalised position $x/L$ along the channel without (**a-b**) and with (**c-d**) applied pressure for $c = 300$ mM. The dashed vertical lines segregate the channel interior, $x/L \in (-0.5, 0.5)$, from the left ($x/L < -0.5$) and right ($x/L > 0.5$) reservoirs. The curves are coloured according to the applied voltage from -50 mV (blue) to 50 mV (orange). **a,** The high friction (BN-like) configuration with $\Delta P/L = 0$. **b,** The low friction (graphite-like) behaviour with $\Delta P/L = 0$. **c,** The high friction (BN-like) configuration with $\Delta P/L = 30$ mbar/μm. **d,** The low friction (graphite-like) behaviour with $\Delta P/L = 30$ mbar/μm.

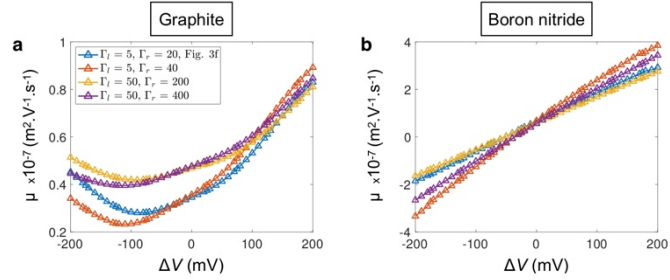

**Extended Data Fig. 9 | Effect of the asymmetry of the system.** $\mu(\Delta V)$ versus $\Delta V$ as a function of asymmetry for: **a,** Low friction (graphite-like) behaviour. In this plot we take $c = 100$ mM, $\alpha_+ = 1$, $\alpha_- = 0.7$, $\mu_+ = \mu_+^{bulk}$, $\mu_- = 0.5\mu_-^{bulk}$ and $\lambda_0/h_0 = 10^{11}$ kg m$^{-3}$ s$^{-1}$, as in the manuscript, while varying the geometric parameters $\Gamma_l$ and $\Gamma_r$, as indicated in the legend. **b,** High friction (BN-like) behaviour, $c = 100$ mM, $\alpha_+ = 0.01$, $\alpha_- = 0.01$, $\mu_+ = \mu_+^{bulk}$, $\mu_- = 0.5\mu_-^{bulk}$ and $\lambda_0/h_0 = 10^{13}$ kg m$^{-3}$ s$^{-1}$, as in the manuscript, while varying the geometric parameters $\Gamma_l$ and $\Gamma_r$, as indicated in the legend in **a**.

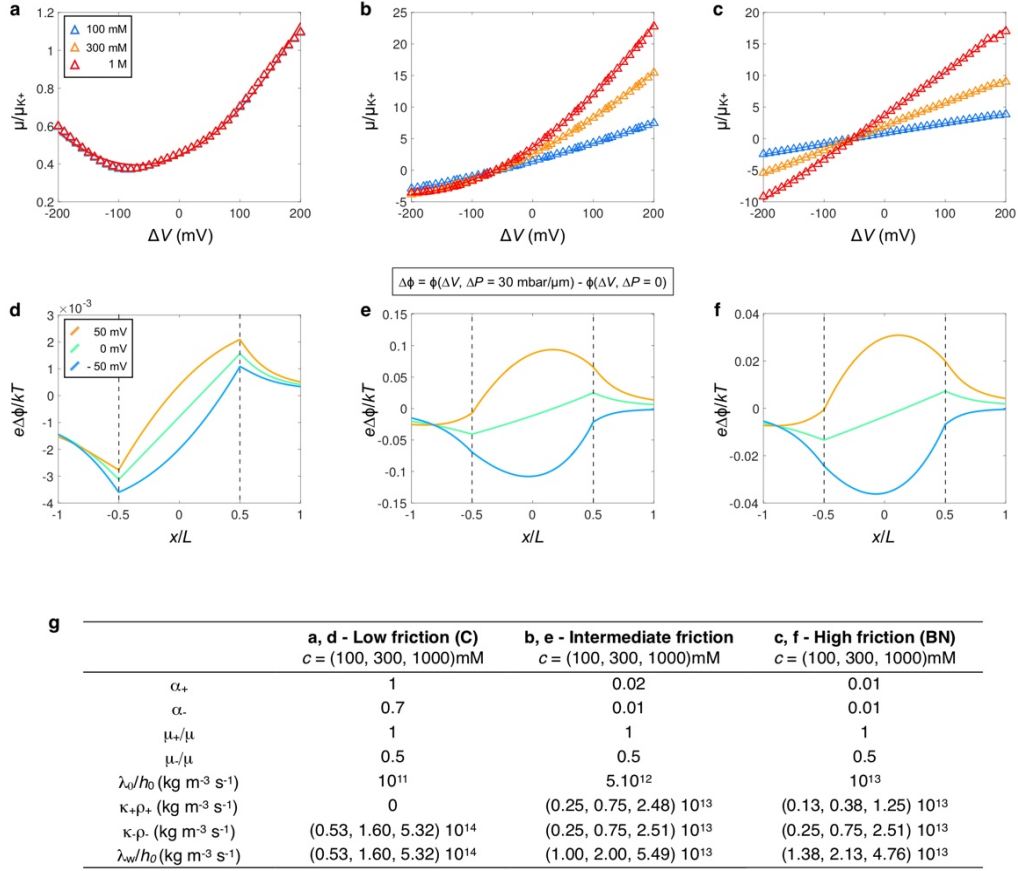

**Extended Data Fig. 10 | Influence of the friction parameters on the model predictions. a-c**, $\mu(\Delta V)$ versus $\Delta V$ for different concentrations ($c$ = 100, 300 and 1,000 mM) and frictional parameters: **a,** Low friction (graphite-like) behaviour. In this plot, we take $\alpha_+ = 1$, $\alpha_- = 0.7$, $\mu_+ = \mu_+^{bulk}$, $\mu_- = 0.5\mu_-^{bulk}$ and $\lambda_0/h_0 = 10^{11}$ kg m$^{-3}$ s$^{-1}$. **b,** Intermediate friction behaviour, $\alpha_+ = 0.02$, $\alpha_- = 0.01$, $\mu_+ = \mu_+^{bulk}$, $\mu_- = 0.5\mu_-^{bulk}$ and $\lambda_0/h_0 = 5.10^{12}$ kg m$^{-3}$ s$^{-1}$. **c,** High friction (BN-like) behaviour, $\alpha_+ = 0.01$, $\alpha_- = 0.01$, $\mu_+ = \mu_+^{bulk}$, $\mu_- = 0.5\mu_-^{bulk}$ and $\lambda_0/h_0 = 10^{13}$ kg m$^{-3}$ s$^{-1}$. **d-f,** Pressure induced variation of the normalised electric potential $\Delta\phi = \phi(\Delta V, \Delta P = 30$ mbar/μm$) - \phi(\Delta V, \Delta P = 0)$ plotted as a function of the normalised channel coordinate $x/L$ axis for $\Delta V$ = -50, 0 and 50 mV. The dashed vertical lines segregate the channel interior, $x/L \in (-0.5, 0.5)$, from the left ($x/L < -0.5$) and right ($x/L > 0.5$) reservoirs. The curves are coloured according to the applied voltage from -50 mV (blue) to 50 mV (orange). Panels **d** through **f** corresponds to the parameters of **a** through **c**, respectively. **g**, Table of the friction parameters corresponding to the data shown in **a** through **c**. The table also shows the decomposition of $\lambda_w(c)$ into its three components for the concentrations considered here.